\documentstyle[twocolumn,prb,aps,epsf]{revtex}

\begin{document}

\draft
\title{
Elementary vortex pinning potential in a chiral $p$-wave superconductor
}

\author{
Nobuhiko Hayashi$^{1}$
and
Yusuke Kato$^{2}$
}

\address{
$^{1}$Computer Center,
Okayama University, Okayama 700-8530, Japan\\
$^{2}$Department of Basic Science,
University of Tokyo, Tokyo 153-8902, Japan
}

\date{\today}

\maketitle

\begin{abstract}
   The elementary vortex pinning potential is studied
in a chiral $p$-wave superconductor with a pairing
${\bf d}=\bar{\bf z}(\bar{k}_{x} \pm i \bar{k}_{y})$
on the basis of the quasiclassical theory of superconductivity.
   An analytical investigation and numerical results
are presented to show that
the vortex pinning potential is dependent on
whether
the vorticity and chirality are parallel or antiparallel.
   Mutual cancellation of
the vorticity and chirality around a vortex
is physically crucial to the effect of the pinning center
inside the vortex core.
\end{abstract}

\pacs{PACS numbers: 74.60.Ge, 74.60.Ec, 74.70.Pq, 74.70.Tx}

   Much attention has been focused on the vortex pinning
in type-II superconductors.
   The vortex pinning
plays an important role on various vortex-related quantities and phenomena
such as the critical current and
the hysteresis of the magnetization
in superconductors under magnetic fields.
   The characteristics of the vortex-related phenomena
are of particular interest in unconventional
superconductors with multiple components of
the superconducting order parameter.\cite{Shung}
   In such superconductors,
multiple states of superconducting order can coexist.
   Accordingly there appear multiple kinds of vortex structure,
where the nature of the vortex pinning can be dependent on
the microscopics of the superconducting order.

   One of the superconductivity with multiple components
of the order parameter
is the chiral $p$-wave one,
${\bf d}=\bar{\bf z}({\bar k}_x \pm i{\bar k}_y)$,
which is composed of two degenerate pairing states
${\bar k}_x$ and ${\bar k}_y$
and breaks the time-reversal symmetry.
   This superfluid $^3$He-A type of chiral $p$-wave pairing state
has been anticipated in
a layered ruthenate superconductor
Sr$_2$RuO$_4$.\cite{Sigrist99}
   While the identification of the genuine superconducting pairing
of this material is still open to
further discussion,\cite{Hasegawa,Izawa,Zhitomirsky,lebed00}
that chiral $p$-wave pairing has
the simplest and essential form and has attracted
a great deal of attention.
  The vortices for Sr$_2$RuO$_4$
have been investigated intensively.\cite{Volovik,Agterberg,Heeb99,Kita,Shiraishi,Matsumoto99,Matsumoto00,Kato00,Goryo,Tewordt,Matsumoto01,Kato01,Takigawa,Nakai}
   In the context of the vortex pinning,
we will see a rich physics contained in that chiral $p$-wave pairing
state.

   In this paper, we investigate the elementary vortex pinning
potential in the chiral $p$-wave superconductor with the pairing
${\bf d}=\bar{\bf z}(\bar{k}_{x} \pm i \bar{k}_{y})$.
   A point-like pinning center and
a single vortex with vorticity perpendicular to conduction layers
in a layered superconductor are considered.
   We show that the vortex pinning potential
depends on the sense of the chirality of the Cooper pairs
relative to the vorticity of the vortex
in the chiral $p$-wave superconductor.
   First we analytically discuss the interplay between
the chirality and vorticity
to explain the mechanism of the chirality dependence
of the vortex pinning potential.
   We then present numerical result for the vortex pinning potential
obtained from self-consistent order parameters.
   Our numerical result confirms the analytical one.
   The chirality dependence of the vortex pinning
would have influence on the hysteresis of the magnetization
and the distribution of the magnetic field in samples,
which might be observed with SQUID and
magneto-optical imaging techniques.

   To investigate the vortex pinning, we use
the quasiclassical theory of superconductivity.\cite{serene}
   We start with the Eilenberger equation for
the quasiclassical Green function
in the absence of the pinning,
%
\begin{equation}
{\hat g}_{\rm imt}(i\omega_n,{\bf r},{\bar{\bf k}})=
-i\pi
\pmatrix{
g_{\rm imt} &
if_{\rm imt} \cr
-if^{\dagger}_{\rm imt} &
-g_{\rm imt} \cr
},
\label{eq:qcg}
\end{equation}
namely,
%
\begin{eqnarray}
i v_{\rm F} {\bar{\bf k}} \cdot
{\bf \nabla}{\hat g}_{\rm imt}
+ \bigl[ i\omega_n {\hat \tau}_{z}-{\hat \Delta},
{\hat g}_{\rm imt} \bigr]
=0,
\label{eq:eilen}
\end{eqnarray}
where
the order parameter is
${\hat \Delta}({\bf r},{\bar{\bf k}}) =
\bigl[ ({\hat \tau}_{x} + i {\hat \tau}_{y}) \Delta({\bf r},{\bar{\bf k}})
- ({\hat \tau}_{x} - i {\hat \tau}_{y}) \Delta^*({\bf r},{\bar{\bf k}})
\bigr] /2$
and ${\hat \tau}_{i}$ the Pauli matrices.
The Eilenberger equation (\ref{eq:eilen})
is supplemented by the normalization condition
%
$
{\hat g}_{\rm imt}(i\omega_n,{\bf r},{\bar{\bf k}})^2
=-\pi^2{\hat 1}
$,
and the commutator is
$[{\hat a},{\hat b}]={\hat a}{\hat b}-{\hat b}{\hat a}$.
   The vector ${\bf r}=(r\cos\phi,r\sin\phi)$
is the center of mass coordinate,
and the unit vector
${\bar{\bf k}}=(\cos\theta,\sin\theta)$
represents
the relative coordinate of the Cooper pair.
   The cylindrical Fermi surface is assumed.
   We use units in which $\hbar = k_{\rm B} = 1$.

   Following
Thuneberg {\it et al.},\cite{Thuneberg82,Thuneberg84,Thuneberg81}
the effect of the pinning is introduced to
the quasiclassical theory of superconductivity
as follows.
   The quasiclassical Green function ${\hat g}$
in the presence of a point-like non-magnetic defect situated at
${\bf r}={\bf R}$
is obtained from the Eilenberger equation
\begin{eqnarray}
i v_{\rm F} {\bar{\bf k}} \cdot
{\bf \nabla}{\hat g}
+ \bigl[ i\omega_n {\hat \tau}_{z}-{\hat \Delta}, {\hat g} \bigr]
=\bigl[ {\hat t}, {\hat g}_{\rm imt} \bigr] \delta ({\bf r}'),
\label{eq:eilen-pin}
\end{eqnarray}
and the $t$ matrix due to the defect
\begin{eqnarray}
{\hat t}(i\omega_n, {\bf r}') =
\frac{v}{D} \Bigl[
{\hat 1} + N_{0} v
\langle {\hat g}_{\rm imt}(i\omega_n, {\bf r}',{\bar {\bf k}}) \rangle_\theta
\Bigr],
\label{eq:t-matrix}
\end{eqnarray}
where ${\bf r}'={\bf r}-{\bf R}$,
the denominator $D=1+(\pi N_0 v)^2 \bigl[
\langle g_{\rm imt} \rangle_\theta ^2
+ \langle f_{\rm imt} \rangle_\theta
\langle f^{\dagger}_{\rm imt} \rangle_\theta
\bigr]$,
the average over the Fermi surface
$\langle \cdots \rangle_\theta = \int \cdots d \theta/2\pi$,
the normal-state density of states on the Fermi surface $N_{0}$,
and
we assume the $s$-wave scattering $v$
when obtaining Eq.\ (\ref{eq:t-matrix}).
   We define a parameter
$\sigma = (\pi N_0 v)^2 /
\bigl[ 1+ (\pi N_0 v)^2 \bigr]$,
which measures how strong
the scattering potential of
the defect is.

   The free energy in the presence of the defect is,
at the temperature $T$, given as
\cite{Thuneberg82,Thuneberg84,Thuneberg81,Viljas}
\begin{eqnarray}
\delta \Omega ({\bf R}) =
N_0 T \int^{1}_{0} d \lambda
\sum_{\omega_{n}}
\int d {\bar {\bf k}}
\int d {\bf r}
{\rm Tr}
\bigl[
\delta {\hat g}_{\lambda} {\hat \Delta}_b
\bigr],
\label{eq:free-ene}
\end{eqnarray}
where $\delta {\hat g}_{\lambda} = {\hat g}-{\hat g}_{\rm imt}$
is evaluated at
${\hat \Delta}= \lambda {\hat \Delta}_{b}$,
and
${\hat \Delta}_{b}$
is the order parameter in the absence of the defect.
Equation (\ref{eq:free-ene}) represents the difference
in the free energy
between the states with and without the defect, and then
gives the vortex pinning potential $\delta \Omega ({\bf R})$.

   For the chiral $p$-wave pairing state
${\bf d}=\bar{\bf z}({\bar k}_x + i{\bar k}_y)
= \bar{\bf z} \exp(i \theta)$,
it is known that
the order parameter around a single vortex,
$\Delta_b({\bf r},{\bar {\bf k}})$
$\bigl[\equiv \Delta_b(r,\phi;\theta) \bigr]$,
has two possible forms
depending on whether
the chirality and vorticity are parallel or
antiparallel each other.\cite{Heeb99,Matsumoto01,Kato01}
%
   One form is
\begin{eqnarray}
\Delta_b^{+-}(r,\phi;\theta)=
\Delta_{+}(r) e^{i(\theta-\phi)}
+ \Delta_{-}(r) e^{i(-\theta+\phi)},
\label{eq:op-pm}
\end{eqnarray}
where the chirality and vorticity are antiparallel (Case I).
   The other is
\begin{eqnarray}
\Delta_b^{++}(r,\phi;\theta)=
\Delta_{+}(r) e^{i(\theta+\phi)}
+ \Delta_{-}(r) e^{i(-\theta+3\phi)},
\label{eq:op-pp}
\end{eqnarray}
where the chirality and vorticity are parallel (Case II).
   Here, the vortex center is situated at ${\bf r}=0$,
the dominant component
$\Delta_{+}(r \rightarrow \infty)=\Delta_{\rm BCS}(T)$,
and
the induced one
$\Delta_{-}(r \rightarrow \infty)=0$.
   Because of axisymmetry of the system,
we can take
$\Delta_{\pm}(r)$
to be real.

   First we analytically investigate the vortex pinning potential.
   We discuss the quantity $\delta \Omega(R=0)$, where
both the defect and the vortex center are situated just at the origin
${\bf r}=0$ ($R \equiv |{\bf R}|$).
   From the quasiclassical viewpoint,
the quasiparticles inside the vortex core,
subject to the Andreev reflection,
run along straight lines called as
quasiparticle paths.\cite{Klein89,Rainer,Hayashi97}
   We consider the quasiparticle paths
which go through the origin ${\bf r}=0$.
   On those paths with zero impact parameter,
the position vector is parallel to the direction of the quasiparticle path
(i.e., ${\bf r} \parallel {\bar {\bf k}}$), and therefore
$\phi=\theta,\ \theta+\pi$.
   In this situation ($\phi=\theta$),
from Eqs.\ (\ref{eq:op-pm}) and (\ref{eq:op-pp}),
the order parameter on the path is
\begin{eqnarray}
\Delta_b^{+-}(r, \phi=\theta; \theta) =
\Delta_{+}(r) + \Delta_{-}(r)
\label{eq:op-pm-c}
\end{eqnarray}
in case I,
and
\begin{eqnarray}
\Delta_b^{++}(r, \phi=\theta; \theta) =
\bigl[ \Delta_{+}(r) + \Delta_{-}(r) \bigr] e^{2i\theta}
\label{eq:op-pp-c}
\end{eqnarray}
in case II.
   The cancellation between
the chirality and vorticity occurs in Eq.\ (\ref{eq:op-pm-c})
and not in Eq.\ (\ref{eq:op-pp-c}).
   Of importance is the resultant difference in the phase factor
of these order parameters.

   On the basis of an analysis of
the so-called zero-core vortex model
in Ref.\ \onlinecite{Thuneberg84},
the matrix elements of
${\hat g}_{\rm imt}$ at the vortex center are
approximately obtained as\cite{z-core}
\begin{equation}
g_{\rm imt} = \sqrt{\omega_n^2 +|{\tilde \Delta}|^2}
\omega_n^{-1}, \quad
f_{\rm imt} = -{\tilde \Delta} \omega_n^{-1}, \quad
f^{\dagger}_{\rm imt} = {\tilde \Delta}^{*} \omega_n^{-1},
\label{eq:z-core-gf}
\end{equation}
where
${\tilde \Delta}=
\Delta_b^{+\pm}(r \rightarrow \infty, \phi=\theta; \theta)$.
   Here, Eq.\ (\ref{eq:z-core-gf}) is obtained with assuming
that
the amplitude of the order parameter
is constant (i.e., zero core) around the vortex,
which is the only approximation in this analysis.
   Inserting the order parameter of Eq.\ (\ref{eq:op-pm-c})
into Eq.\ (\ref{eq:z-core-gf}),
we obtain
the anomalous Green functions
integrated over the Fermi surface as, in case I,
\begin{equation}
\langle f_{\rm imt} \rangle_\theta = f_{\rm imt}, \quad
\langle f_{\rm imt}^{\dagger} \rangle_\theta = f_{\rm imt}^{\dagger}
\label{eq:f-caseI}
\end{equation}
because of the absence of any phase factors in Eq.\ (\ref{eq:op-pm-c}),
i.e., because of the cancellation between the chirality factor
$\exp(i \theta)$
and the vorticity factor
$\exp(-i \phi)$
in Eq.\ (\ref{eq:op-pm}).
On the other hand, in case II,
\begin{equation}
\langle f_{\rm imt} \rangle_\theta = 0, \quad
\langle f_{\rm imt}^{\dagger} \rangle_\theta = 0
\label{eq:f-caseII}
\end{equation}
because of the phase factor $\exp(2i\theta)$ contained in
Eq.\ (\ref{eq:op-pp-c}).
   The diagonal component of $\langle {\hat g}_{\rm imt} \rangle_\theta$
is $\langle g_{\rm imt} \rangle_\theta = g_{\rm imt}$
both in cases I and II.
   Consequently, in case I,
$\langle {\hat g}_{\rm imt} \rangle_\theta  =  {\hat g}_{\rm imt}$
and we obtain
$[{\hat t},{\hat g}_{\rm imt}]=0$ from Eq.\ (\ref{eq:t-matrix}).
   In case II,
$\langle {\hat g}_{\rm imt} \rangle_\theta  \neq  {\hat g}_{\rm imt}$
and
$[{\hat t},{\hat g}_{\rm imt}] \neq 0$ generally.

   When $[{\hat t},{\hat g}_{\rm imt}]=0$,
the Eilenberger equation (\ref{eq:eilen-pin})
in the presence of the defect
is identical to Eq.\ (\ref{eq:eilen})
(the equation in the absence of the defect),
namely, the defect has no influence on the Green function and
the free energy.
   From this and the above results of the analysis
of the factor $[{\hat t},{\hat g}_{\rm imt}]$,
we find that
$\delta \Omega(0) = 0$ in case I
when the chirality is antiparallel to the vorticity,
and $\delta \Omega(0) \neq 0$ in case II
when the sense of the chirality is the same as that of the vorticity.
   It means that the vortex pinning depends on the chirality
in the chiral $p$-wave superconductor.

   The above analytical result is based on the zero-core vortex model,
i.e., on the non-self-consistent (constant) amplitude of
the order parameter.
   We next investigate the vortex pinning potential $\delta \Omega (R)$
numerically with the self-consistent order parameters around the vortex
which have the forms of Eqs.\ (\ref{eq:op-pm}) and (\ref{eq:op-pp}).
   As self-consistent
amplitude $\Delta_{\pm}(r)$ in Eqs.\ (\ref{eq:op-pm})
and (\ref{eq:op-pp}),
we adopt numerical data which we have obtained in Ref.\ \onlinecite{Kato01}
by solving
self-consistently the Eilenberger equation.

\begin{figure}
\epsfxsize=75mm
\begin{center}
\epsfbox{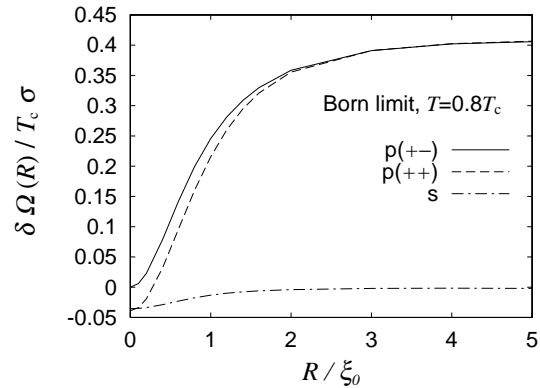}
\end{center}
%
     \vspace{-4mm}
(a)
\epsfxsize=75mm
\begin{center}
\epsfbox{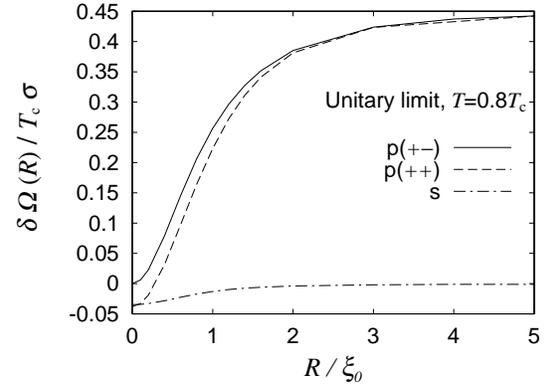}
\end{center}
%
     \vspace{-4mm}
(b)
\epsfxsize=75mm
\begin{center}
\epsfbox{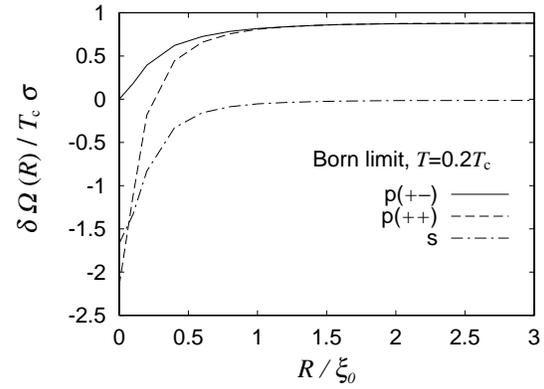}
\end{center}
%
     \vspace{-4mm}
(c)
\epsfxsize=75mm
\begin{center}
\epsfbox{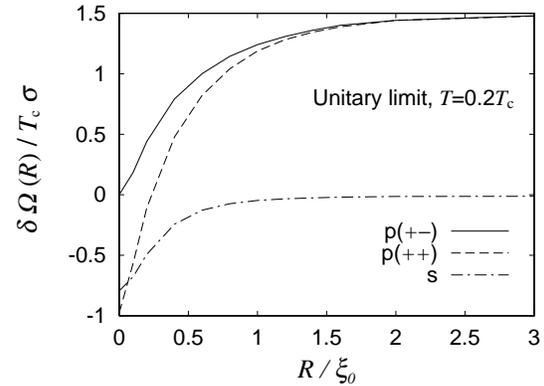}
\end{center}
    \vspace{-4mm}
(d)
\vspace{5.5mm}
%
%
\caption{
   The vortex pinning potential as a function of
the distance $R$ between the vortex center and the defect.
   Solid lines $\bigl[$p($+-$)$\bigr]$
correspond to the case of the $p$-wave pairing
with the chirality antiparallel to the vorticity (Case I).
   Dashed lines $\bigl[$p($++$)$\bigr]$
correspond to the case of the $p$-wave pairing
with the chirality parallel to the vorticity (Case II).
   Dot-dashed lines correspond to the case of
the isotropic $s$-wave pairing.
   $T_{\rm c}$ is the superconducting critical temperature.
   The distance $R$ is normalized with
the coherence length $\xi_0 = v_{\rm F} / \Delta_{\rm BCS}(T=0)$.
}
\label{fig:1}
\end{figure}

   In Fig.\ \ref{fig:1},
we show the numerical results for
$\delta \Omega (R)$
in the Born limit ($\sigma \ll 1$)
and the unitary limit ($\sigma \rightarrow 1$).
   We present those results for the chiral $p$-wave pairing and
the isotropic $s$-wave one.
   As noted in Fig.\ \ref{fig:1},
in the case of the $s$-wave pairing (dot-dashed lines),
the difference in the free energy
between the states with and without the defect,
$\delta \Omega (R)$,
is equal to zero at $R \rightarrow \infty$
($R$ is the distance between the vortex center and the defect).
   This is because the Anderson's theorem \cite{Anderson59}
is valid far away from the vortex core.
   On the other hand, in the chiral $p$-wave pairing cases
(solid and dashed lines),
$\delta \Omega (R \rightarrow \infty)$ is finite and positive
as seen in Fig.\ \ref{fig:1}.
   The quantity $\delta \Omega (R \rightarrow \infty)$
is equal to the loss of the condensation energy
due to the pair breaking effect of
the defect far away from the vortex core
(i.e., the breakdown of the Anderson's theorem).
   As noted in Figs.\ \ref{fig:1}(a) and \ref{fig:1}(b),
at $T=0.8T_{\rm c}$ (high temperature),
the condensation energy loss in bulk
$\delta \Omega (R \rightarrow \infty)$
dominantly
contributes to
the depth of the vortex pinning potential $\delta \Omega (R)$,
i.e., to the vortex pinning energy.
   From Figs.\ \ref{fig:1}(a) and \ref{fig:1}(b)
it is noticed that
   the vortex pinning energies of the chiral $p$-wave pairing cases
at a high temperature
are
about 10 times larger than those of the $s$-wave pairing case.
   This enhancement of the pinning effect
is due to the breakdown of the Anderson's theorem,
and then it must be a common feature of unconventional
superconductors.\cite{Thuneberg81,Friesen,Kulic,Hayashi}
   For example, in the case of high-$T_{\rm c}$ cuprates,
this may be one of the reasons why small
point defects such as Zn atoms\cite{Pan} and oxygen vacancies\cite{Kes}
are efficient pinning centers.

   As noted in Figs.\ \ref{fig:1}(c) and \ref{fig:1}(d),
at $T=0.2T_{\rm c}$ (low temperature),
the contribution of the vortex core ($R \simeq 0$)
to the depth of $\delta \Omega (R)$
is nonzero in case II (dashed lines).
   Here, the contribution of the vortex core
means the energy gain due to the presence of the
scattering center in the vortex core.
   In contrast,
the depth of $\delta \Omega (R)$,
i.e., the vortex pinning energy, is determined in case I (solid lines)
only by the loss of the condensation energy far away from the vortex core.
   It is noticeable that certainly
$\delta \Omega (R=0)$ equals to zero in case I.
   This numerical result confirms the analytical one discussed above.
   The vortex pinning energy depends on
whether the chirality and vorticity are antiparallel (solid lines)
or parallel (dashed lines).
   Especially in the Born limit,
the difference in the vortex pinning energy is eminent
as noticed in Fig.\ \ref{fig:1}(c),
because in this limit the loss of the condensation energy in bulk
is relatively small compared to
the contribution of the vortex core to
the depth of $\delta \Omega (R)$.

   In general, the two chiral states of cases I and II can coexist
as domain structure in samples under magnetic fields.
   The spatial gradient of the magnetic field in a sample
is proportional to the local strength of the vortex pinning
in the critical state.
   In terms of the present chirality-dependent vortex pinning,
the gradient inside the domain of the case-II state
is predicted to be steeper than that inside the domain of the case-I state.
   This may be experimentally observed as a signature of the chiral state.
   Also the domain structure of the two chiral states
depends on the hysteresis of applied magnetic field,
and therefore the present chirality-dependent vortex pinning
may affect the hysteresis curve
during multiple cycles of the magnetization
as observed\cite{Shung} in UPt$_3$.

   In the case of the usual winding-1 vortex
$\Delta \propto \exp(i\phi)$,
the chiral ^^ ^^ $p$-wave" pairing
$\bar{k}_{x} \pm i \bar{k}_{y} = \exp(\pm i \theta)$
is essential for the cancellation
between the chirality and vorticity.
If winding-2 vortices
$\Delta \propto \exp(2 i\phi)$
are realized in a chiral $d$-wave state
$\bar{k}_x^2 - \bar{k}_y^2 \pm i \bar{k}_x \bar{k}_y = \exp(\pm 2 i \theta)$,
the same kind of cancellation occurs.

   We comment on the relation of the present vortex pinning
to the superconducting gap structure in Sr$_2$RuO$_4$.
   In this material,
it has been pointed out from experiments that
the gap has line nodes\cite{Hasegawa,Izawa,Zhitomirsky}
and little in-plane anisotropy.\cite{Izawa}
   Models for the gap structure consistent with those experimental facts
were proposed, in which there existed
horizontal line nodes perpendicular to the axis of the cylindrical
Fermi surface.\cite{Izawa,Zhitomirsky}
   Now, for the present theory of
the chirality-dependent vortex pinning,
what is important is that
the Fermi surface averages of the anomalous Green functions
(i.e., the average of the order parameter except for the chiral part)
are finite as in Eq.\ (\ref{eq:f-caseI}).
   The present chirality dependence of the vortex pinning
does not occur,
if the order parameters have sign changes on all Fermi surfaces
relevant to superconductivity
as
$\Delta({\bf k})\sim \exp(\pm i\theta)\cos(ck_z)$.
   It occurs, if there are no sign changes
as
$\Delta({\bf k})\sim \exp(\pm i\theta)|\cos(ck_z)|$.
   In another case,\cite{Zhitomirsky}
the chirality dependence is expected to occur
when the order parameter is nodeless on
the major Fermi surface with dominant density of states,
even if there are gap nodes and sign changes
on the other minor Fermi surfaces.

   In conclusion,
   we investigated the elementary vortex pinning potential
$\delta \Omega (R)$
on the basis of the quasiclassical theory of superconductivity.
   In the chiral $p$-wave pairing state,
$\delta \Omega (R)$ was dependent on the sense of the chirality
relative to the vorticity at a low temperature.
In terms of the present chirality-dependent vortex pinning,
a theoretical analysis
for anomalies in the hysteresis of the magnetization
observed experimentally in Sr$_2$RuO$_4$\cite{Tamegai}
would be interesting and is left for future work.

   We thank T.\ Tamegai, P.\ H.\ Kes, T.\ Kita,
K.\ Kinoshita, and A.\ Maeda
for helpful discussions,
and J.\ R.\ Clem, E.\ V.\ Thuneberg, and M.\ L.\ Kuli\'c
for considerate correspondence.
   One of the authors (N.H.) also
thanks M.\ Ichioka, K.\ Machida, M.\ Takigawa,
N.\ Nakai, M.\ Matsumoto,
S.\ H.\ Pan, and M. Fogelstr\"om for useful discussions and comments.
   Y.K.\ thanks M.\ Sigrist for useful discussions.
   This work is partly supported by
Grant-in-Aid for Scientific Research on Priority Areas (A)
of ^^ ^^ Novel Quantum Phenomena in Transition Metal Oxides" (12046225)
from the Ministry of Education, Science, Sports and Culture and
Grant-in-Aid for Encouragement of Young Scientists from Japan Society for
the Promotion of Science (12740203).

%
%
\vspace{-0.6cm}

%
%
\vspace{-0.5cm}
%
%
\end{document}